\documentclass [letter,twocolumn]{jpsj3}

\usepackage{txfonts}
\usepackage{color}
\usepackage{ulem}
\usepackage{graphicx}% Include figure files
\usepackage{dcolumn}% Align table columns on decimal point
\usepackage{bm}% bold math

\title{
Pressure-Induced Superconductivity from Doping-Induced Antiferromagnetic Phase of 112-type Ca$_{1-x}$La$_{x}$FeAs$_{2}$
}

\author{
Shunsaku~Kitagawa$^{1}$\thanks{Present address: Department of Physics, Kyoto University, Kyoto 606-8502, Japan}\thanks{E-mail address: kitagawa.shunsaku.8u@kyoto-u.ac.jp}, 
Taishi~Sekiya$^{1}$, 
Yo~Fujiyoshi$^{1}$, 
Shingo~Araki$^{1}$, 
Tatsuo~C.~Kobayashi$^{1}$, 
Naoki~Nishimoto$^{2}$, 
Tasuku~Mizukami$^{2}$, 
Satoshi~Ioka$^{2}$, 
Kazunori~Fujimura$^{2}$, 
Kazutaka~Kudo$^{2}$, and
Minoru~Nohara$^{2}$
}
\inst{
$^1$Department of Physics, Okayama University, Okayama 700-8530, Japan \\
$^2$Research Institute for Interdisciplinary Science, Okayama University, Okayama 700-8530, Japan
}

\date{\today}

\abst{
The effects of pressure on antiferromagnetic (AFM) and superconducting phase transitions of 112-type Ca$_{1-x}$La$_{x}$FeAs$_{2}$ were studied, and the in-plane electrical resistivity $\rho_{ab}$ was measured with an indenter-type pressure cell. 
The AFM phase transition temperatures of $T_{\rm N}$ = 47, 63, and 63 K at ambient pressure for $x$ = 0.18, 0.21, and 0.26 was suppressed by applying pressure $P$, with superconductivity emerging at critical pressures of $P_{\rm c}$ $\simeq$ 0, 1.5, and 3.4 GPa, respectively. 
Correspondingly, the slope of $T_{\rm N}$ against $P$ decreased as $dT_{\rm N}/P$ $\simeq$ $-$15 and $-$2 K/GPa for $x$ = 0.21 and 0.26, respectively. 
Thus, although the AFM phase was stabilized with La doping $x$, the AFM phase was suppressed by pressure, and superconductivity eventually emerged. 
}

\begin{document}
\clearpage
\maketitle

The parent compounds of most iron-based superconductors exhibit antiferromagnetic (AFM) ordering, and superconductivity emerges with high transition temperature, $T_{\rm c}$, when the AFM phase is suppressed by pressure or chemical doping. \cite{J.Paglione_Naturephys_2010}
Some iron-based superconductors, however, exhibit AFM ordering that is induced by pressure or chemical substitution. 
For example, in hydrogen-doped LaFeAsO$_{1-x}$H$_{x}$, the substitution of H$^{-}$ for O$^{2-}$ results in the suppression of superconductivity, and an AFM phase emerges at $x \geq 0.4$, where the electrons are overdoped. \cite{M.Hiraishi_NatPhys_2014,N.Fujiwara_PRL_2013}
In phosphorous-doped LaFeAs$_{1-x}$P$_{x}$O, an AFM phase emerges at $0.4 \leq x \leq 0.7$ because of the substitution of P$^{3-}$ for As$^{3-}$. \cite{S.Kitagawa_JPSJ_2014,K.T.Lai_PRB_2014,T.Shiota_JPSJ_2016,S.Miyasaka_PRB_2017}
Phosphorus doping also results in the emergence of an AFM phase in (Ca$_4$Al$_2$O$_6$)Fe$_2$(As$_{1-x}$P$_{x}$)$_2$ at $0.3 \leq x \leq 0.95$. \cite{H.Kinouchi_PRB_2013}
In these cases of phosphorus doping, no charge carriers are introduced because phosphorus and arsenic are isovalent elements. 
A pressure-induced AFM phase has been reported in FeSe at $P \geq 0.8$~GPa. \cite{M.Bendele_PRL_2010,M.Bendele_PRB_2012,K.Kothapalli_NatCommun_2016,P.S.Wang_PRL_2016}
Interestingly, in FeSe, the application of higher pressures suppresses the AFM phase, and the superconducting transition temperature $T_{\rm c}$ exhibits a sudden increase up to 38 K at 6 GPa. \cite{J.P.Sun_NatCommun_2016}

Very recently, a novel iron-based superconductor exhibiting doping-induced AFM ordering has been discovered in the 112-type compound Ca$_{1-x}$La$_{x}$FeAs$_{2}$. \cite{N.Katayama_JPSJ_2013,K.Kudo_JPSJ_2014,K.Kudo_JPSJ_2014-2,H.Yakita_JACS_2014,M.Nohara_APX_2017}
The Ca$_{1-x}$La$_{x}$FeAs$_{2}$ compound crystallizes into a monoclinic structure with space group $P$2$_1$ (No. 4, $C_{2}^{2}$) 
with alternately stacked FeAs and Ca$_{1-x}$La$_{x}$As layers along the $c$-axis.
While the end member, or the parent compound without La doping CaFeAs$_{2}$, was not obtained, the substitution of La for Ca stabilized the 112 phase at $0.15 \leq x \leq 0.27$ for Ca$_{1-x}$La$_{x}$FeAs$_{2}$.
The highest value of $T_{\rm c}$ = 35 K has been found at $x = 0.15$ in Ca$_{1-x}$La$_{x}$FeAs$_{2}$.\cite{K.Kudo_JPSJ_2014}
Substitution of Sb$^{3-}$ for As$^{3-}$ further enhanced $T_{\rm c}$ up to 47 K in Ca$_{1-x}$La$_{x}$Fe(As$_{1-y}$Sb$_{y}$)$_{2}$. \cite{K.Kudo_JPSJ_2014-2,Ota_JPSJ_2017}
Moreover, $^{75}$As nuclear magnetic resonance (NMR) measurements revealed that the substitution of La$^{3+}$ for Ca$^{2+}$ in Ca$_{1-x}$La$_{x}$FeAs$_{2}$ suppresses superconductivity and results in AFM ordering with a N\'eel temperature of $T_{\rm N}$ = 70 K for $x$ = 0.24, where the electrons are thought to be overdoped. \cite{S.Kawasaki_PRB_2015} 
Subsequently, neutron diffraction measurements of Ca$_{0.73}$La$_{0.27}$FeAs$_{2}$ revealed a stripe-type AFM ordering at $T_{\rm N}$ = 54 K, following a monoclinic-to-triclinic structural phase transition at $T_{\rm s}$ = 58 K. \cite{S.Jiang_PRB_2016}
Interestingly, the AFM ordering induced by La-doping is suppressed by the doping of either Co or Ni, 
and superconductivity emerges in Ca$_{0.74(1)}$La$_{0.26(1)}$Fe$_{1-y}$Co$_{y}$As$_{2}$ at $y$ $>$ 0.02~\cite{S.Jiang_PRB_2016_2} and in Ca$_{1-x}$La$_{x}$Fe$_{1-y}$Ni$_{y}$As$_{2}$ at $y$ $>$ 0.004 and $x$ = 0.18 and 0.24. \cite{T.Xie_arXiv_2017}
Although both La and Co/Ni doping produce electron charge carriers, La and Co/Ni doping have opposing effects in Ca$_{1-x}$La$_{x}$FeAs$_{2}$ materials: La doping suppresses superconductivity and induces AFM ordering, while Co/Ni doping suppresses AFM ordering and induces superconductivity. 
This raises the question of whether the application of pressure enhances AFM ordering, as in the case of La doping, or suppresses AFM ordering, as in the case of Co/Ni doping.

In this paper, we present the results of resistivity measurements under various pressures in Ca$_{1-x}$La$_{x}$FeAs$_{2}$ with $x$ = 0.18, 0.21, and 0.26, which exhibited AFM ordering at $T_{\rm N}$ = 47, 63, and 63 K, respectively. 
Our results revealed that the application of pressure suppressed the AFM ordering and induced superconductivity. 
Thus, the AFM ordering of Ca$_{1-x}$La$_{x}$FeAs$_{2}$ exhibited a ``normal'' response to pressure.

%\section{Experimental}
Single crystals of Ca$_{1-x}$La$_{x}$FeAs$_{2}$ were grown by heating a mixture of Ca, La, FeAs, and As powders as described elsewhere. \cite{K.Kudo_JPSJ_2014-2} 
The La content $x$ was analyzed by energy-dispersive X-ray spectrometry. 
The in-plane electrical resistivity $\rho_{ab}$ was measured with a standard four-probe method.
Pressure was generated in an indenter-type pressure cell, \cite{T.C.Kobayashi_RSI_2007}
and we used Daphne 7474 as pressure-transmitting medium. \cite{K.Murata_RSI_2008}
The applied pressure was determined from the superconducting transition temperature $T_{\rm c}$ of the lead manometer as $P = [T_{\rm c}(0) - T_{\rm c}(P)]/0.364$ in GPa. \cite{A.Eiling_JPFMP_1981,B.Bireckoven_JPESI_1988}

%%%%%%%%%%%%%%%%%%%%%%%%%%% Figure 1 %%%%%%%%%%%%%%%%%%%%%%%%%%%%%%%%%%%%%
\begin{figure*}[!tb]
\vspace*{0mm}
\begin{center}
\includegraphics[width=15cm,clip]{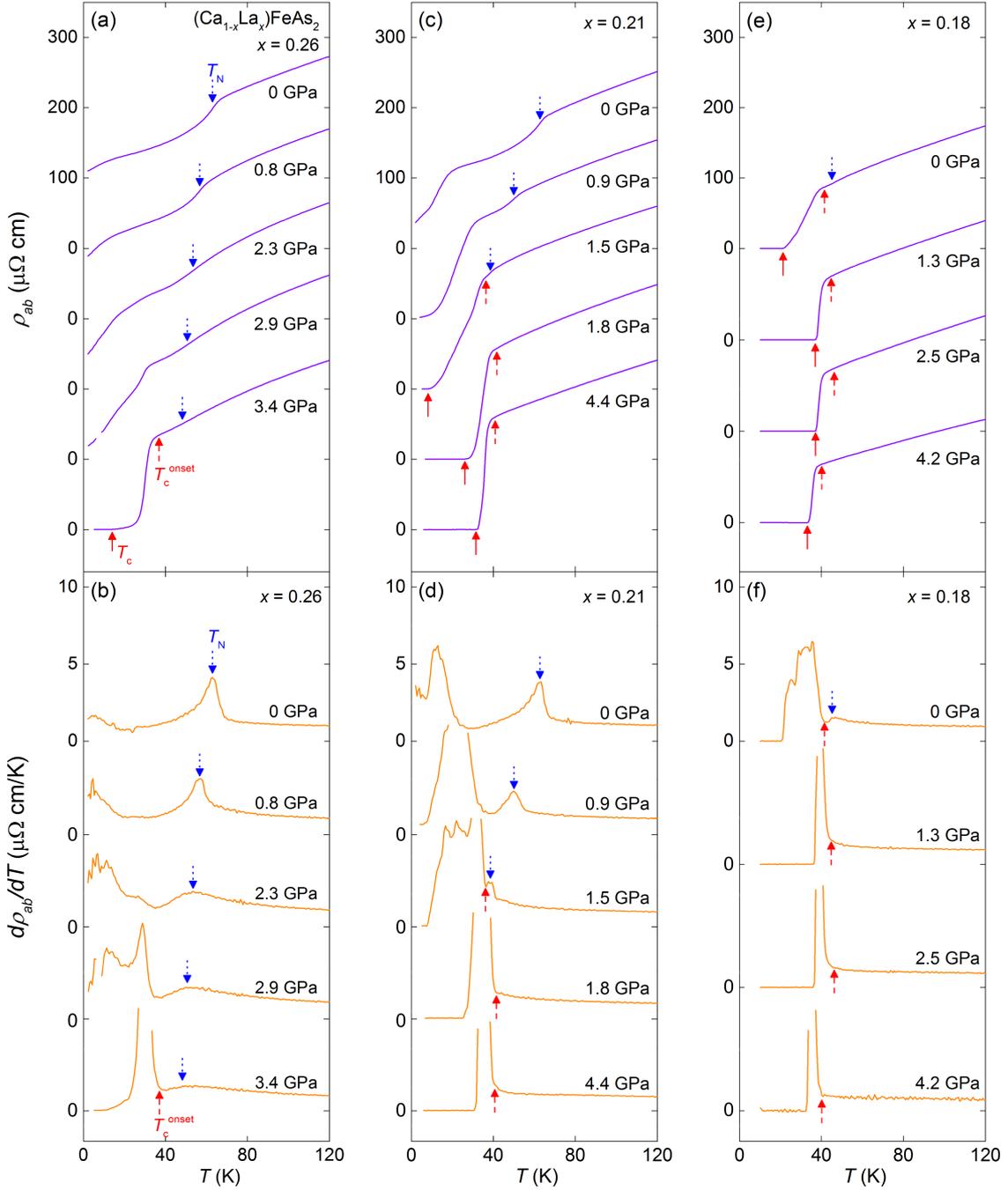}
\end{center}
\vspace*{00mm}
\caption{
(Color online) Temperature dependence of $\rho_{ab}$ and its temperature derivative $d\rho_{ab}/dT$ under various pressures in (Ca$_{1-x}$La$_{x}$)FeAs$_{2}$ with $x$ = 0.26, 0.21, and 0.18. 
The blue dotted arrows indicate the antiferromagnetic transition temperature $T_{\rm N}$ determined by a peak in $d\rho_{ab}/dT$.
The red solid and broken arrows indicate the superconducting transition temperature $T_{\rm c}$, where zero resistivity was observed, and the onset temperature $T_{\rm c}^{\rm onset}$, where $d\rho_{ab}/dT$ exhibited a steep increase upon cooling, respectively.
}
\label{Fig.1}
\end{figure*}

%%%%%%%%%%%%%%%%%%%%%%%%%%%%%%%%%%%%%%%%%%%%%%%%%%%%%%%%%%%%%%%%%%%%%%%%%%%

%%%%%%%%%%%%%%%%%%%%%%%%%%% Figure 2 %%%%%%%%%%%%%%%%%%%%%%%%%%%%%%%%%%%%%
\begin{figure}[!tb]
\vspace*{0mm}
\begin{center}
\includegraphics[width=5.5cm,clip]{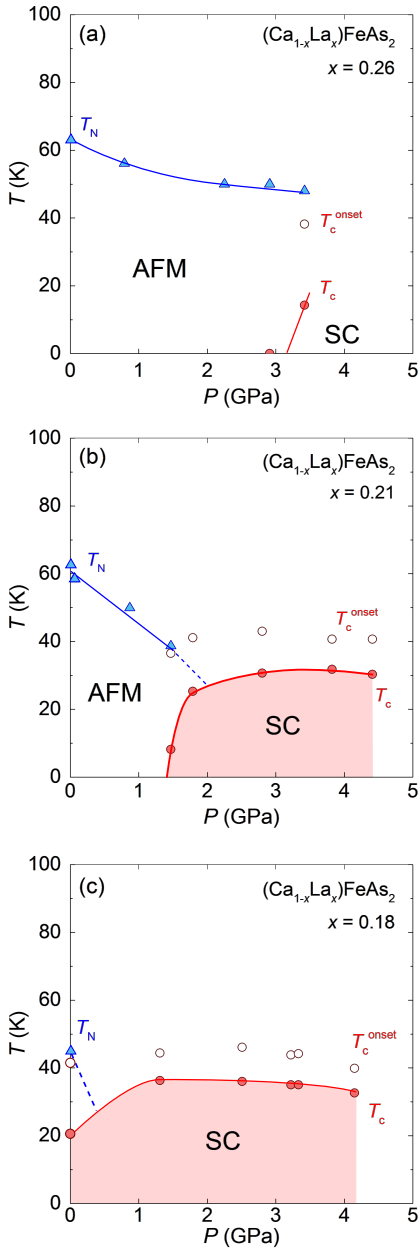}
\end{center}
\vspace*{00mm}
\caption{
(Color online) Pressure--temperature phase diagrams determined by resistivity measurements for (Ca$_{1-x}$La$_{x}$)FeAs$_{2}$ with $x$ = 0.26, 0.21, and 0.18. 
Here, AFM and SC denote the antiferromagnetic and superconducting phases, respectively. 
Triangles represent the antiferromagnetic transition temperature $T_{\rm N}$, while filled and open circles represent the superconducting transition temperature $T_{\rm c}$, where zero resistivity was observed, and the onset temperature $T_{\rm c}^{\rm onset}$, where $d\rho_{ab}/dT$ exhibited a steep increase upon cooling, respectively.
}
\label{Fig.2}
\end{figure}

%%%%%%%%%%%%%%%%%%%%%%%%%%%%%%%%%%%%%%%%%%%%%%%%%%%%%%%%%%%%%%%%%%%%%%%%%%%

Figures 1(a)--1(f) show the temperature dependence of $\rho_{ab}$ and its temperature derivative $d\rho_{ab}/dT$ as a function of temperature $T$ under different pressures in (Ca$_{1-x}$La$_{x}$)FeAs$_{2}$ with $x$ = 0.18, 0.21, and 0.26.
At the highest doping $x = 0.26$ and ambient pressure, $\rho_{ab}$ showed a kink at approximately 65 K, which became more evident in the $d\rho_{ab}/dT$ vs. $T$ data, which had a peak at 63 K, as shown in Fig. 1(b).
Previous neutron diffraction and transport measurements at $x = 0.27$ confirmed that the peak in $d\rho_{ab}/dT$ was belonging to the AFM transition temperature $T_{\rm N}$. \cite{S.Jiang_PRB_2016}
Accordingly, we determined $T_{\rm N}$ = 63 K from the peak temperature of $d\rho_{ab}/dT$.
Compared to temperature values in previous reports, this transition temperature was consistent with, but slightly different from, $T_{\rm N}$ = 70 K, which was determined by NMR measurements for $x = 0.24$, \cite{S.Kawasaki_PRB_2015} and $T_{\rm N}$ = 54 K, which was determined by neutron diffraction measurements for $x$ = 0.27. \cite{S.Jiang_PRB_2016} 
The monoclinic-to-triclinic structural phase transition temperature $T_{\rm s}$ cannot be obtained from the $d\rho_{ab}/dT$ vs. $T$ curve, which may have a kink at $T_{\rm s}$ at a slightly higher temperature than $T_{\rm N}$, as reported by Jiang $et~al$. \cite{S.Jiang_PRB_2016}

As shown in Figs. 1(a) and 1(b), $T_{\rm N}$ gradually decreased with the increasing pressure.
The peak in $d\rho_{ab}/dT$ became considerably broader above 2.3 GPa, but it was still existing up to 3.4 GPa, which was the maximum pressure measured. 
At 3.4 GPa, we found a superconducting transition at $T_{\rm c} = 14$ K, which was determined from zero resistivity and is indicated by a red arrow in Fig. 1(a). 
This is the first example of pressure-induced superconductivity in a 112 system.
We determined an onset temperature $T_{\rm c}^{\rm onset}$, where $d\rho_{ab}/dT$ exhibited a steep increase with the decreasing temperature, as indicated by a broken red arrow in Fig. 1(b). 
While it was unclear whether the AFM phase microscopically coexisted with superconductivity in the present resistivity measurements, based on magnetic susceptibility and $\mu$SR measurements, the microscopic coexistence of superconductivity and antiferromagnetism in Ca$_{0.74(1)}$La$_{0.26(1)}$Fe$_{1-y}$Co$_{y}$As$_{2}$ was reported. \cite{S.Jiang_PRB_2016}

Thus, the determined $T_{\rm N}$, $T_{\rm c}$, and $T_{\rm c}^{\rm onset}$ values are summarized in the $P$--$T$ phase diagram for Ca$_{1-x}$La$_{x}$FeAs$_{2}$ with $x$ = 0.26, as shown in Fig. 2(a). 
The sample exhibited an AFM ordering at $T_{\rm N}$ = 63 K at ambient pressure. 
The transition temperature $T_{\rm N}$ decreased with the increasing pressure $P$ by an initial slope of $dT_{\rm N}/dT$ $\simeq$ $-$8.6 K/GPa. 
At higher pressures, the slope decreased with $dT_{\rm N}/dT$ $\simeq$ $-$2 K/GPa, suggesting that the AFM phase of $x$ = 0.26 was robust against pressure, while $T_{\rm N}$ decreased to 49 K at 3.4 GPa. 
Superconductivity at $T_{\rm c}$ = 14 K suddenly emerged at this pressure, which we defined as the critical pressure $P_{\rm c}$ (= 3.4 GPa). 

In the same manner, we obtained pressure--temperature phase diagrams for $x$ = 0.21 and 0.18, as shown in Figs. 2(b) and 2(c), respectively. 
The $x$ = 0.21 sample exhibited an AFM transition at $T_{\rm N}$ = 63 K, which was almost the same as that observed for $x$ = 0.26, at ambient pressure. 
However, the transition temperature $T_{\rm N}$ rapidly decreased with the increasing pressure by a slope of $dT_{\rm N}/dP$ $\simeq$ $-$15 K/GPa, which was noticeably steeper than that observed for  $x$ = 0.26. 
Correspondingly, superconductivity emerged at $P_{\rm c}$ = 1.5 GPa, and $T_{\rm c}$ approached a maximum value of 32 K at 3.8 GPa. 
It was not clear whether the AFM phase coexisted with superconductivity at $P$ $>$ $P_{\rm c}$ (= 1.5 GPa) in the present resistivity measurements, because the onset of the superconducting transition at $T_{\rm c}^{\rm onset}$ prevented the observation of any anomalies that can be associated with AFM transition at $T_{\rm N}$ ($<$ $T_{\rm c}^{\rm onset}$). 
For the $x$ = 0.18 sample, 
the AFM phase was reduced to $T_{\rm N}$ = 47 K, and superconductivity emerged at $T_{\rm c}$ = 20.5 K at ambient pressure. 
A maximum of $T_{\rm c}$ = 36.3 K was observed at 1.3 GPa, which was consistent with a previous report. \cite{W.Zhou_EPL_2015} 
The AFM transition was not observed at $P$ $>$ 1.3 GPa, as  observed from the monotonic temperature dependence of $\rho_{ab}$ in the normal state shown in Fig. 1(e).

Our experimental results clearly demonstrate that the application of pressure suppressed the doping-induced AFM ordering, and resulted in a superconducting phase.
Reasonable nesting, which is essential to the AFM ordering, was observed between the Fermi surfaces at the $\Gamma$ and M points by angle-resolved photoemission spectroscopy measurements in Ca$_{0.73}$La$_{0.27}$FeAs$_{2}$. \cite{S.Jiang_PRB_2016}
Thus, we believe that the application of pressure weakened the Fermi surface nesting and resulted in superconductivity.

%\section{Conclusion}
The pressure dependence of $T_{\rm N}$ and $T_{\rm c}$ determined by in-plane resistivity $\rho_{ab}$ has been investigated in Ca$_{1-x}$La$_{x}$FeAs$_{2}$ ($x = 0.18, 0.21$, and 0.26) in order to understand the effect of pressure for doping-induced AFM ordering.
When pressure was applied, $T_{\rm N}$ decreased, and superconductivity appeared after the suppression of AFM ordering.
This is the first example of pressure-induced superconductivity in a 112 system.
The critical pressure $P_{\rm c}$, defined as the pressure where zero resistivity is observed, was $\simeq$ 3.4, 1.5, and 0 GPa for $x$ = 0.26, 0.21, and 0.18, respectively.
The effect of pressure for the AFM ordering and superconductivity is opposite to that of La doping, but similar to those of Co and Ni doping.

\section*{Acknowledgments}
Part of this work was performed at the Advanced Science Research Center, Okayama University. 
This work was partially supported by Grants-in-Aid for Scientific Research (Nos. JP15H01047, JP23244075, JP25400372, and JP26287082) provided by the Japan Society for the Promotion of Science (JSPS) and the Program for Advancing Strategic International Networks to Accelerate the Circulation of Talented Researchers from JSPS.


\begin{thebibliography}{10}

\bibitem{J.Paglione_Naturephys_2010}
J.~Paglione and R.~L. Greene, Nat. Phys. {\bfseries 6}, 645 (2010).

\bibitem{M.Hiraishi_NatPhys_2014}
M.~Hiraishi, S.~Iimura, K.~M. Kojima, J.~Yamaura, H.~Hiraka, K.~Ikeda, P.~Miao, Y.~Ishikawa, S.~Torii, M.~Miyazaki, I.~Yamauchi, A.~Koda, K.~Ishii, M.~Yoshida, J.~Mizuki, R.~Kadono, R.~Kumai, T.~Kamiyama, T.~Otomo, Y.~Murakami, S.~Matsuishi, and H.~Hosono, 
Nat. Phys. {\bfseries 10}, 300 (2014).

\bibitem{N.Fujiwara_PRL_2013}
N.~Fujiwara, S.~Tsutsumi, S.~Iimura, S.~Matsuishi, H.~Hosono, Y.~Yamakawa, and H.~Kontani, 
Phys. Rev. Lett. {\bfseries 111}, 097002 (2013).

\bibitem{S.Kitagawa_JPSJ_2014}
S.~Kitagawa, T.~Iye, Y.~Nakai, K.~Ishida, C.~Wang, G.-H. Cao, and Z.-A. Xu, 
J. Phys. Soc. Jpn. {\bfseries 83}, 023707 (2014).

\bibitem{K.T.Lai_PRB_2014}
K.~T. Lai, A.~Takemori, S.~Miyasaka, F.~Engetsu, H.~Mukuda, and S.~Tajima, 
Phys. Rev. B {\bfseries 90}, 064504 (2014).

\bibitem{T.Shiota_JPSJ_2016}
T.~Shiota, H.~Mukuda, M.~Uekubo, F.~Engetsu, M.~Yashima, Y.~Kitaoka, K.~T. Lai, H.~Usui, K.~Kuroki, S.~Miyasaka, and S.~Tajima, 
J. Phys. Soc. Jpn. {\bfseries 85}, 053706 (2016) .

\bibitem{S.Miyasaka_PRB_2017}
S. Miyasaka, M. Uekubo, H. Tsuji, M. Nakajima, S. Tajima, T. Shiota, H. Mukuda, H. Sagayama, H. Nakao, R. Kumai, and Y. Murakami, 
Phys. Rev. B {\bfseries 95}, 214515 (2017).

\bibitem{H.Kinouchi_PRB_2013}
H.~Kinouchi, H.~Mukuda, Y.~Kitaoka, P.~M. Shirage, H.~Fujihisa, Y.~Gotoh, H.~Eisaki, and A.~Iyo, 
Phys. Rev. B {\bfseries 87}, 121101(R) (2013).

\bibitem{M.Bendele_PRL_2010}
M.~Bendele, A.~Amato, K.~Conder, M.~Elender, H.~Keller, H.-H. Klauss, H.~Luetkens, E.~Pomjakushina, A.~Raselli, and R.~Khasanov, 
Phys. Rev. Lett. {\bfseries 104}, 087003 (2010).

\bibitem{M.Bendele_PRB_2012}
M. Bendele, A. Ichsanow, Yu. Pashkevich, L. Keller, Th. Str\"assle, A. Gusev, E. Pomjakushina, K. Conder, R. Khasanov, and H. Keller, 
Phys. Rev. B {\bfseries 85}, 064517 (2012). 

\bibitem{K.Kothapalli_NatCommun_2016}
K.~Kothapalli, A.~E. Bohmer, W.~T. Jayasekara, B.~G. Ueland, P.~Das, A.~Sapkota, V.~Taufour, Y.~Xiao, E.~Alp, S.~L. Bud'ko, P.~C. Canfield, A.~Kreyssig, and A.~I. Goldman, 
Nat. Commun. {\bfseries 7}, 12728 (2016).

\bibitem{P.S.Wang_PRL_2016}
P.~S. Wang, S.~S. Sun, Y.~Cui, W.~H. Song, T.~R. Li, R.~Yu, H.~Lei, and W.~Yu, 
Phys. Rev. Lett. {\bfseries 117},  237001 (2016).

\bibitem{J.P.Sun_NatCommun_2016}
J.~P. Sun, K.~Matsuura, G.~Z. Ye, Y.~Mizukami, M.~Shimozawa, K.~Matsubayashi, M.~Yamashita, T.~Watashige, S.~Kasahara, Y.~Matsuda, J.~Q. Yan, B.~C. Sales, Y.~Uwatoko, J.~G. Cheng, and T.~Shibauchi, 
Nat. Commun. {\bfseries 7}, 12146 (2016).

\bibitem{N.Katayama_JPSJ_2013}
N.~Katayama, K.~Kudo, S.~Onari, T.~Mizukami, K.~Sugawara, Y.~Sugiyama, Y.~Kitahama, K.~Iba, K.~Fujimura, N.~Nishimoto, M.~Nohara, and H.~Sawa, 
J. Phys. Soc. Jpn. {\bfseries 82}, 123702 (2013).

\bibitem{K.Kudo_JPSJ_2014}
K.~Kudo, T.~Mizukami, Y.~Kitahama, D.~Mitsuoka, K.~Iba, K.~Fujimura, N.~Nishimoto, Y.~Hiraoka, and M.~Nohara, 
J. Phys. Soc. Jpn. {\bfseries 83}, 025001 (2014).

\bibitem{K.Kudo_JPSJ_2014-2}
K.~Kudo, Y.~Kitahama, K.~Fujimura, T.~Mizukami, H.~Ota, and M.~Nohara, 
 J. Phys. Soc. Jpn. {\bfseries 83} (2014) 093705.

\bibitem{H.Yakita_JACS_2014}
H.~Yakita, H.~Ogino, T.~Okada, A.~Yamamoto, K.~Kishio, T.~Tohei, Y.~Ikuhara, Y.~Gotoh, H.~Fujihisa, K.~Kataoka, H.~Eisaki, and J. Shimoyama, 
J. Am. Chem. Soc. {\bfseries 136}, 846 (2014).

\bibitem{M.Nohara_APX_2017}
M.~Nohara and K.~Kudo, 
Advances in Physics: X {\bfseries 2}, 450 (2017). 

\bibitem{Ota_JPSJ_2017}
H. Ota, K. Kudo, T. Kimura, Y. Kitahama, T. Mizukami, S. Ioka, and M. Nohara, 
J. Phys. Soc. Jpn. {\bfseries 86},  025002 (2017). 

\bibitem{S.Kawasaki_PRB_2015}
S.~Kawasaki, T.~Mabuchi, S.~Maeda, T.~Adachi, T.~Mizukami, K.~Kudo, M.~Nohara, and G.-q.~Zheng, 
Phys. Rev. B {\bfseries 92}, 180508(R) (2015) .

\bibitem{S.Jiang_PRB_2016}
S.~Jiang, C.~Liu, H.~Cao, T.~Birol, J.~M. Allred, W.~Tian, L.~Liu, K.~Cho, M.~J. Krogstad, J.~Ma, K.~M. Taddei, M.~A. Tanatar, M.~Hoesch, R.~Prozorov, S.~Rosenkranz, Y.~J. Uemura, G.~Kotliar, and N.~Ni, 
Phys. Rev. B {\bfseries 93}, 054522 (2016).

\bibitem{S.Jiang_PRB_2016_2}
S.~Jiang, L.~Liu, M.~Sch\"utt, A.~M. Hallas, B.~Shen, W.~Tian, E.~Emmanouilidou, A.~Shi, G.~M. Luke, Y.~J. Uemura, R.~M. Fernandes, and N.~Ni, 
Phys. Rev. B {\bfseries 93}, 174513 (2016).

\bibitem{T.Xie_arXiv_2017}
T.~Xie, D.~Gong, W.~Zhang, Y.~Gu, Z.~Husges, D.~Chen, Y.~Liu, L.~Hao, S.~Meng, Z.~Lu, S.~Li, and H.~Luo, 
Supercond. Sci. Technol. {\bfseries 30}, 095002 (2017).

\bibitem{T.C.Kobayashi_RSI_2007}
T.~C. Kobayashi, H.~Hidaka, H.~Kotegawa, K.~Fujiwara, and M.~I. Eremets, 
Rev. Sci. Instrum. {\bfseries 78},  023909 (2007).

\bibitem{K.Murata_RSI_2008}
K.~Murata, K.~Yokogawa, H.~Yoshino, S.~Klotz, P.~Munsch, A.~Irizawa, M.~Nishiyama, K.~Iizuka, T.~Nanba, T.~Okada, Y.~Shiraga, and S.~Aoyama, 
Rev. Sci. Instrum. {\bfseries 79},  085101 (2008).

\bibitem{A.Eiling_JPFMP_1981}
A.~Eiling and J.~S. Schilling, J. Phys. F: Met. Phys. {\bfseries 11}, 623 (1981).

\bibitem{B.Bireckoven_JPESI_1988}
B.~Bireckoven and J.~Wittig, J. Phys. E: Sci. Instrum. {\bfseries 21}, 841 (1988).

\bibitem{W.Zhou_EPL_2015}
W.~Zhou, X.~Z. Xing, X.~Zhou, M.~X. Xu, and Z.~X. Shi, Europhys. Lett. {\bfseries 109}, 37005 (2015).

\end{thebibliography}
\end{document}